\documentclass[10pt,final, journal, twocolumn]{IEEEtran}

\ifCLASSINFOpdf

\else

\fi

\usepackage{booktabs}
\usepackage{graphicx}
\usepackage{booktabs}
\usepackage{diagbox}
\usepackage{multirow}
\usepackage{caption}

\usepackage{array}
\usepackage{algorithm}
\usepackage{cite}
\usepackage{amssymb}
\usepackage{algpseudocode}
\usepackage{color}
\usepackage{amsmath}
\usepackage{subfig}
\usepackage{stfloats}

\makeatletter 
\let\myorg@bibitem\bibitem
\def\bibitem#1#2\par{%
	\@ifundefined{bibitem@#1}{%
		\myorg@bibitem{#1}#2\par
	}{%
		\begingroup
		\color{\csname bibitem@#1\endcsname}%
		\myorg@bibitem{#1}#2\par
		\endgroup
	}%
}

\makeatother 

\begin{document}
\captionsetup[figure]{name={Fig.},labelsep=period}
\captionsetup{font={small}}
\rmfamily


\title{\fontsize{16pt}{20pt}{Betweenness Centrality Based Dynamic Source Routing for Flying Ad~Hoc Networks in Marching Formation}}

\author{Shaoshi~Yang,~\IEEEmembership{Senior Member,~IEEE,} Wei Zhao, Chu-Meng Wang, Wen-Yu~Dong, Xiaojie Ju 
\thanks{
Copyright \copyright 2025 IEEE. Personal use of this material is permitted. However, permission to use this material for any other purposes must be obtained from the IEEE by sending a request to pubs-permissions@ieee.org.
	
This work was supported in part by the Siyuan Artificial Intelligence Science and Technology Collaborative Innovation Alliance under Grant SY240101203 and in part by the Beijing Municipal Natural Science Foundation under Grant L242013. (\textit{Corresponding author: Shaoshi Yang.})

S. Yang, W. Zhao, C.-M. Wang, and W.-Y. Dong are with the School of Information and Communication Engineering, Beijing University of Posts and Telecommunications, with the Key Laboratory of Universal Wireless Communications, Ministry of Education, and also with the Key Laboratory of Mathematics and Information Networks, Ministry of Education, Beijing 100876, China (e-mail: \{shaoshi.yang, wei.zhao, wangchumeng99, wenyu.dong\}@bupt.edu.cn). 

X. Ju is with China Academy of Launch Vehicle Technology, Beijing 100076, China (e-mail: juxiaojie789@163.com).
}
}

\maketitle
\begin{abstract}
Designing high-performance routing protocols for flying ad hoc networks (FANETs) is challenging due to the diversity of applications and the dynamics of network topology. The existing general-purpose routing protocols for ad hoc networks often oversimplify mobility patterns and disregard the unequal importance of nodes, resulting in suboptimal routing decisions that are unsuitable for task-oriented FANETs. To break the bottleneck, in
this paper we propose a betweenness centrality based dynamic source routing (BC-DSR) protocol for a flying ad hoc network (FANET) in marching formation. Firstly, we introduce a Gauss-Markov group (GMG) mobility model based on the leader-follower pattern, which accurately captures the temporal and spatial correlations of node movements in the realistic marching formation. Besides,
we exploit the concept of BC defined in graph theory to measure the structural unequal importance of relay nodes, i.e., to determine link weights, in the particular marching formation topology. The path of least cost is calculated relying on a weighted directed graph constructed. 
The ns-3 based simulation results demonstrate that our BC-DSR protocol achieves higher packet-delivery ratio and lower average end-to-end latency and routing overhead ratio than representative benchmark protocols used in FANETs, while maintaining a reasonably small network jitter.
\end{abstract}

\begin{IEEEkeywords}
Betweenness centrality, flying ad hoc network, mobility model, routing protocol, unmanned flying platforms.
\end{IEEEkeywords}

\IEEEpeerreviewmaketitle
\section{Introduction}
\IEEEPARstart{I}{n} the beyond-5G and 6G era, non-terrestrial networks (NTNs) \cite{NTN_6G} relying on unmanned flying platforms (UFPs) have stimulated substantial interests in numerous areas, such as military operation, disaster rescue, emergency communication, and transportation traffic monitoring~\cite{2016Survey}. Typical UFPs include unmanned aerial vehicles (UAVs), balloons, low/medium/high altitude platforms, and the low/medium/high earth orbit satellites. To take advantage of UFPs, the concept of flying ad hoc network (FANET) composed of multiple collaborative UFPs is advocated, where routing protocols are of central importance. The dynamic network topology and frequent link disruptions make routing requirements of FANETs more challenging than those of the conventional mobile ad hoc networks (MANETs) or vehicular ad hoc networks (VANETs) \cite{lakewrouting}.

In recent years, some efforts have been devoted to improving the classical routing protocols to adapt to FANETs, including topology-based and position-based approches. For example, a position-based modified optimized link state routing protocol (PM-OLSR) \cite{gangopadhyayposition} improves network throughput and reduces latency by leveraging 3D position information, residual energy and node degree to optimize multipoint relays (MPRs) selection. Besides, a cross-layer and energy-aware on-demand distance vector (CLEA-AODV) \cite{mansourcross} routing protocol enhances energy efficiency and routing stability, while a dynamic source routing protocol based on path reliability and link monitoring repair mechanism (DSR-PM) \cite{liangdynamic} ensures communication stability and reliability. Hosseinzadeh et al. \cite{hosseinzadehnew} improved greedy perimeter stateless routing (GPSR) by limiting flooding during the greedy forwarding process. There are also a few cluster-based schemes \cite{khedrmwcrsf, khayatredundant} that capture the structural advantages of cluster heads in preserving network connectivity and facilitating inter-cluster data transmission. In addition to these classical methods, intelligent optimization algorithms have emerged as a promising solution to addressing the unique challenges of FANETs. For instance, Wei et al. \cite{QFAGR} proposed a Q-learning-based fast adaptive geographic routing (QFAGR) protocol that optimizes routing decisions by adjusting parameters based on local topology changes, improving throughput and reducing delays. However, the above-mentioned protocols tend to oversimplify mobility patterns and overlook the structural characteristics of the formation, offering limited insights into optimizing routing decisions for task-oriented FANETs. Moreover, intelligent routing protocols also face challenges like poor generalization ability, high complexity, and slow convergence, which can hinder their effectiveness in highly dynamic or large-scale FANETs.

\indent As demonstrated in \cite{kimfanet}, the performance of routing protocols in FANETs varies with the mobility models employed, and making the integration of mobility information into routing protocol design is a crucial task. Hu et al. \cite{HuCyberPhysical} proposed a cyber–physical routing protocol exploiting trajectory planning information (CPR-TD) for a mission-oriented FANET, while Zhou et al. \cite{zhouoptimized} introduced an efficient trajectory-based optimized routing protocol (TORP) for UAV swarms. However, the pre-defined deterministic trajectory model’s slow response to network topology changes limits flexibility and scalability while increasing communication and computational overhead.

In this paper, we consider efficient routing protocol design for high-mobility FANETs operating in a marching formation, which exhibits unique characteristics of having both random and deterministic ingredients in the mobility pattern, and thus structural unequal importance of relay nodes. Our main contributions are as follows.
\begin{itemize}
\item We propose a Gauss-Markov group (GMG) mobility model to accurately characterize the marching formation of FANETs. Our model captures both the temporal and spatial correlations of nodes' motion, as well as the moderate randomness of individual movements, altogether.
\item We propose a betweenness centrality based dynamic source routing (BC-DSR) protocol for FANETs. By comparing the betweenness centrality of each node distributedly, the relay nodes that are more reliable can be efficiently selected, resulting in a more robust route. 
\item We conduct ns-3 based simulations and the results show that our BC-DSR protocol achieves higher packet-delivery ratio (PDR), lower average end-to-end (E2E) latency and routing overhead ratio (ROR) than the widely used benchmark routing protocols, while maintaining a reasonably small network jitter.
\end{itemize}
\section{Modeling the FANET in Marching Formation}
\subsection{Network Topology}
\begin{figure}[htbp]
	\centering
	\includegraphics[width=3.5in]{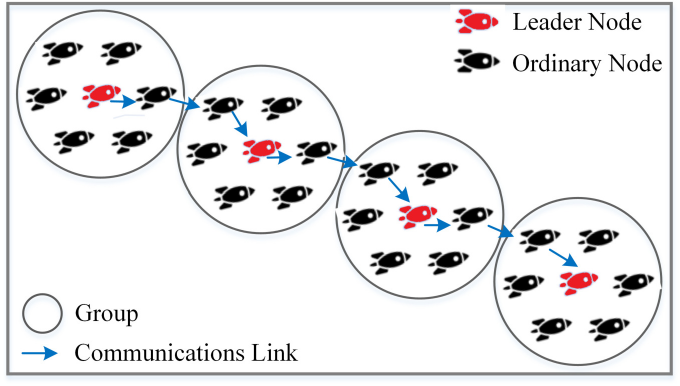}
	\caption{The topology of a FANET in marching formation.}
	\label{Fig1}
\end{figure}

Consider a FANET, where multiple UAV groups are arranged in a long and confined zone and move towards a common destination in a marching formation\footnote {It allows for potential communication disruptions and interference.} to cooperatively perform a task, as shown in Fig.\ref{Fig1}. In each group, a leader is initially pre-defined based on the network topology and task requirements, with the remaining nodes acting as ordinary nodes. However, the leader role is adaptive and can be reassigned based on real-time network conditions. Reassignment occurs when an ordinary node’s performance, evaluated through key metrics\footnote {These metrics include, e.g., normalized spatial proximity, velocity following factor, average link retention rate, and node degree deviation \cite{HACO}.}, surpasses that of the current leader, triggering the role-switching mechanism.
It is assumed, reasonably enough, that the inter-UAV links are mainly line-of-sight (LoS) channels and do not change rapidly during maneuver. The communications between nodes in each group may have to take multiple hops, and in general most communications between nodes belonging to different groups need to be relayed by their respective leader nodes. Obviously, leader nodes are more important than ordinary nodes, and there is an directivity when data is delivered between two nodes in different groups. Therefore, we assume the leader nodes must be selected as part of the forwarding nodes to establish communication links.
\subsection{Mobility Model}
Since the mobility of nodes in the above topology is spatially and temporally correlated, in addition to some randomness, we propose a GMG mobility model\footnote{For a FANET that moves in a marching formation, the motion of each node exhibits the Markov property, i.e., the motion state at a given time instant depends on that of the previous time instant. Meanwhile, there might be moderate randomness associated with individual node movements. The GM mobility model is capable of characterizing both the temporal correlations and the randomness to some degree, while the RPG mobility model is suitable for describing the spatial correlations.} that gleans the benefits of both the reference point group (RPG) mobility model\cite{2015RPGM} and the Gauss-Markov (GM) mobility model \cite{2018An}. Firstly, a leader node is selected as the reference node in each group, and its mobility pattern is characterized by a GM process. Then the ordinary nodes move around the leader nodes. The movement of each leader node is time-dependent, and its speed and direction changes are continuous. More specifically, according to the definition of the GM  mobility model, the speed and the direction of motion of the leader node in each group during a given time interval satisfy:
\begin{equation}\label{eq1}
{v_{t + 1}} = \alpha {v_t} + (1 - \alpha )\bar v + \sqrt {1 - {\alpha ^2}}{\kappa_t},
\end{equation}
\begin{equation}\label{eq2}
{\theta _{t + 1}} = \alpha {\theta _t} + (1 - \alpha )\bar \theta  + \sqrt {1 - {\alpha ^2}}{\rho_t},
\end{equation}
\begin{equation}\label{eqadd3}
{\phi_{t + 1}} = \alpha {\phi _t} + (1 - \alpha )\bar \phi  + \sqrt {1 - {\alpha ^2}}{\zeta_t},
\end{equation}
where ${v_{t + 1}}$, ${\theta _{t + 1}}$ and $\phi_{t + 1}$ represent the speed, azimuth and elevation of the leader node at time instant $t+1$, respectively; $\alpha$ is a constant between $[0,1]$ reflecting the degree of memory; $\bar v$, $\bar \theta$ and $\bar \phi$ represent the leader node's mean speed, mean azimuth and mean elevation when $t\rightarrow \infty$, respectively; $\kappa_t$, $\rho_t$ and $\zeta_t$ are random variables obeying $\mathcal N(0, \sigma_\kappa^2)$, $\mathcal N(0, \sigma_\rho^2)$ and $\mathcal N(0, \sigma_\zeta^2)$, with $\sigma_\kappa$, $\sigma_\rho$ and $\sigma_\zeta$ being the standard deviation of $v_{t}$, $\theta_{t}$ and $\phi_{t}$,   respectively, when $t\rightarrow \infty$. In addition, we have
\begin{equation}\label{eq3}
{x_{t + 1}} = {x_t} + {v_t}  t  \cos {\theta _t} \cos {\phi_t},
\end{equation}
\begin{equation}\label{eq4}
{y_{t + 1}} = {y_t} + {v_t}  t  \sin {\theta _t} \cos {\phi_t},
\end{equation}
\begin{equation}\label{eqadd5}
{z_{t + 1}} = {z_t} + {v_t}  t  \sin {\phi_t},
\end{equation}
where $(x_t, y_t, z_t)$ represents the 3-D coordinates of the leader node at time $t$.

\begin{figure}[tbp]
	\centering
	\includegraphics[width=3.5in]{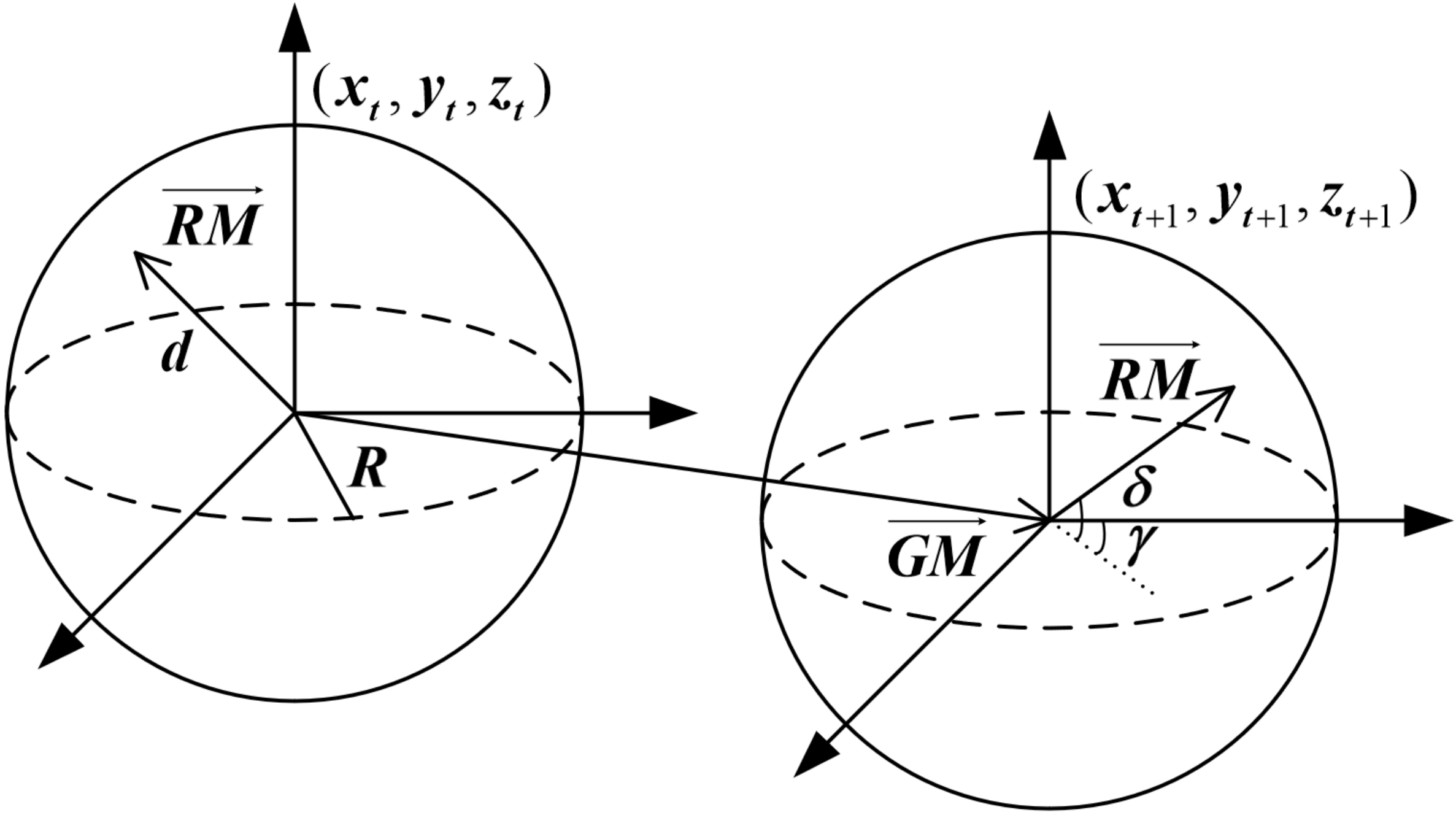}
	\caption{Illustration of the GMG mobility model.}
	\label{Fig2}
\end{figure}

Fig.\ref{Fig2} illustrates how the reference node and the ordinary nodes move from time $t$ to $t+1$. The reference node moves with the group motion vector $\overrightarrow {\textrm{GM}}  = ({x_{t + 1}} - {x_t},{y_{t + 1}} - {y_t}, {z_{t + 1}} - {z_t})$, while in each group the new position of an ordinary node is generated by adding a random motion vector $\overrightarrow {\textrm{RM}}$ to the new position  $({x_{t{\rm{ + }}1}},{y_{t{\rm{ + }}1}},{z_{t{\rm{ + }}1}})$ of the reference node. $\overrightarrow {\textrm{RM}}$ of length $d$ is uniformly distributed within a ball that has a radius $R$ and is centered at the reference node. The azimuth $\gamma$ and elevation $\delta$ of $\overrightarrow {\textrm{RM}}$ are uniformly distributed within $[0^{\circ}, 360^{\circ}]$ and $[-90^{\circ}, 90^{\circ}]$, respectively. Each $\overrightarrow {\textrm{RM}}$  is independent from the corresponding ordinary node’s previous locations. The establishment of the GMG mobility model is to make the nodes move according to the roughly predefined trajectory so as to grasp the location information of nodes. In this way, we can get the connectivity between nodes to construct a series of weighted directed graphs.

\section{The proposed BC-DSR routing protocol}
\subsection{Packet Header}
\begin{figure}[htbp]
	\centering
	\includegraphics[width=2.5in]{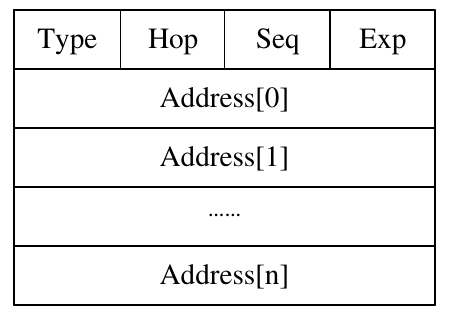}
	\caption{The packet header.}
	\label{Fig3}
\end{figure}

In our routing protocol, the source node stores all the feasible paths in the packet header when it sends a packet. The packet header format is shown in Fig.\ref{Fig3}. We divide all the packets into data packets and control packets. Control packets include route request (RREQ), route reply (RREP) and route error (RERR) packets. In the packet header, we use the Type field to indicate the packet type, which is $8$-bit long. The Hop field also contains $8$ bits and indicates the total number of hops from the source node to the destination node. When sending packets, we assign a $8$-bit sequence number, namely the value in the Seq field, to each packet. The remaining $8$ bits are allocated to the Exp field, which indicates a packet's expiration time, after which the packet is discarded before it has been delivered to the destination node. We use $32$ bits to represent the address of each node on the path. Address[0], Address[1], $\cdots$, Address[n] represent the address of each node from the source node to the destination node, and they constitute the route of a complete packet delivery.
\subsection{Route Establishment}
\begin{figure}[t]
	\centering
	\includegraphics[width=3.5in]{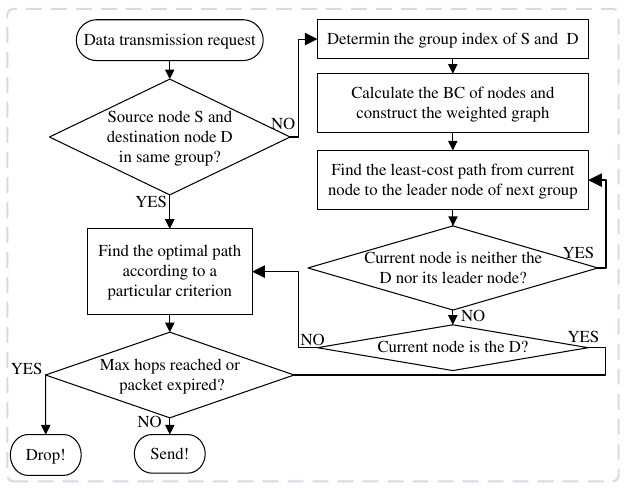}
	\caption{The flowchart of the process of route establishment.}
	\label{Fig4}
\end{figure}

The establishment of a route constitutes two situations: 
1) the source node S and the destination node D belong to the same group; 2) they belong to different groups. The route establishment process starts only when data packets need to be transmitted. Fig.\ref{Fig4} is a flowchart of the route establishment, which illustrates the specific route establishment process. The details are as follows:
\begin{enumerate}
  \item When a data packet needs to be transmitted, we first check whether the source and destination nodes are in the same group. If they are, we use a particular algorithm, such as the Dijkstra algorithm, to find the optimal path between the nodes, based on the predefined weight of each edge.
  \item If the two are in different groups, we first determine the group index of both, and then calculate the BC of nodes, as detailed later.
  \item After getting the BC of each node, we construct a weighted graph where the weight of each edge is given by the reciprocal of the sum of the BC of the two connected nodes.
  \item The least-cost path is iteratively searched from the source node to the leader node of the next group until the leader node of the group containing the destination is reached.
  \item If the leader node is not the destination node, continue to find the optimal path by a particular algorithm according to the predefined weight for each edge.
  \item During data transmission, the sending or receiving node of each hop determines whether the current hop count has exceeds the maximum allowed hops or whether the packet has exceeds its lifetime. If either limit is exceeded, the packet is discarded.
\end{enumerate}
\subsection{Weight of Edge Calculation in BC-DSR}
The weight of each edge invoked in the above route establishment process is calculated by using the concept of BC. In graph theory, BC \cite{Betweenness} mearsures the importance of a node by the frequency of this node lying on the shortest path between two other nodes. The more frequently a node acts as a relay, the greater its centrality is. A large value of BC of a node means that many or all shortest paths between other nodes must pass through it. If this node is dysfunctional, communications between other nodes become difficult and may even suffer an outage (because the originally shortest path breaks). In an $N$-node network modeled by a graph, where the set of vertices and edges are denoted as $\mathbb V$ and $\mathbb E$, respectively, the normalized BC of the node $n$ is calculated as
\begin{equation}\label{eq6}
{C_{\rm B}(n)= \frac{\sum_{i,j,n \in \mathbb V, i\neq n \neq j} \sigma_{i j}(n)/\sigma_{i j}}{(N-1)(N-2)}},
\end{equation}
\noindent where $C_{\rm B}(n) \in [0, 1]$, ${\sigma _{ij}}(n)$ is the number of shortest paths that pass through node $n$ between any two other nodes $i$ and $j$ in the network, and ${{\sigma _{ij}}}$ indicates the number of shortest paths from node $i$ to node $j$. Within the $N$-node network, $N-1$ represents the number of choices for node $i$, and $N-2$ for node j. 

For calculating ${C_{\rm B}}(n)$, let us define
\begin{equation}\label{eq9}
{{\mathbb P}_i}(j) = \{ m \in {\mathbb V}| (m,j)\in \mathbb{E}, {d}(i,j) = {d}(i,m) + \omega (m,j)\},
\end{equation}
where ${{\mathbb P}_i}(j)$ represents the set of precursor nodes for node $j$ in the shortest path from node $i$ to node $j$, while both $d(i,j)$ and $d(i,m)$ denote the cumulative weights (i.e. distance) of the shortest path from a given node to another\footnote{There is only a single hop from $m$ to $j$, hence $\omega(m,j)$ is the weight of the edge $(m,j)$.}. (\ref{eq9}) indicates that the shortest path from $i$ to $j$ must start from $i$ to $m$ and then reach $j$ via the edge $(m,j)$, where $m$ is the precursor node of $j$. Then, the following relationship holds:
\begin{equation}\label{eq10}
{\sigma _{ij}} = \sum\limits_{m \in {P_i}(j)} {{\sigma _{im}}}, \forall i \ne j.
\end{equation}
The classic breadth-first search (BFS) algorithm is then used to calculate ${{\sigma _{ij}}}$. This is because in BFS the nodes are visited according to the distance from node $i$ exactly in the ascending order. The node $m$ is the precursor of node $j$, thus we have $d(i,m)<d(i,j)$. Therefore, when node $j$ is reached, ${{\sigma _{im}}}$ of its precursor node $m$ must have been calculated, then all of ${{\sigma _{ij}}}$ can be calculated. To calculate ${C_{\rm B}}(n)$, we also need to calculate ${{\sigma _{ij}}(n)}$ according to:
\begin{equation}\label{eq11}
{\sigma _{ij}}(n) = \left\{ {\begin{array}{*{10}{c}}
{{\sigma _{in}} \times {\sigma _{nj}},}\\
0,
\end{array}} \right.\begin{array}{*{10}{c}}
\\
\end{array}\begin{array}{*{10}{c}}
{\textrm{if}\; d(i,n) + d(n,j) = d(i,j);}\\
{\textrm{else.}}
\end{array}
\end{equation}
Finally, the weight of an edge $(n_1, n_2)$ is given by
\begin{equation}\label{eq12}
w_{n_1,n_2} = \frac{1}{{C_{\rm B}}(n_1) + {C_{\rm B}}(n_2)},
\end{equation}
which is used by the shortest-path algorithm invoked.
\subsection{Route Recovery}
An RERR messaging mechanism is adopted for route recovery. In this mechanism, when a particular hop of a route suffers an outage, an RERR message is fed back to the source node hop by hop. More specifically, when a node ready to transmit packets detects that the next hop is unable to support a successful transmission, it sends back an RERR message hop by hop along the reverse route until it reaches the source node. In this process, each intermediate node on the reverse route receives the RERR message from its neighbor node. Finally, the source node initiates the route establishment process again and searches for the appropriate route to the destination node.
\section{Performance evaluation}
\subsection{Performance Metrics}
We compare the proposed BC-DSR protocol with  four representative benchmark protocols, including AODV, DSR, DSR-PM and CPR-TD, based on the ns-3 network simulator. The performance metrics considered are as follows.
\begin{enumerate}
  \item Average E2E latency: It characterizes the average time interval between sending and receiving a packet across a network, i.e.,
        \begin{equation}\label{eq13}
         \overline D_{\rm{E2E}}  = \frac{\sum D_{i,\rm{E2E}}}{N_{\rm R}} = \frac{{\sum {(t_{i,\rm R} - t_{i,\rm S})} }}{{{N_{\rm R}} }},
        \end{equation}
where $t_{i,\rm S}$ and $t_{i,\rm R}$ represent the time instant of sending the packet $i$ from a given source node and receiving the packet $i$ at a given destination node, respectively, and $N_{\rm R}$ is the total number of packets received at these destination nodes. Note that such a single time interval comprises both the time consumed by establishing routes and the time occupied by data forwarding.
\begin{table}[tbp]
	\centering
	\caption{Main Simulation Parameters}
	\label{Table 1}
	\begin{tabular}{cc}
		\toprule
		Parameters & Value\\
		\midrule
		Network simulator & ns-3 (release ns-3.27)\\
		Network size & $6 \times 6$, $6\times 12$, $10\times 10$, $12\times 12$\\
		Mobility model & GMG mobility model\\
		Mean speed of leader nodes ($\bar v$) & 200 m/s\\
		$(\alpha, \bar \theta, \bar \phi)$ in the GMG model & $(0.5,\ 0^\circ,\ 0^\circ)$ \\
		$(\sigma_\kappa,\sigma_\rho,\sigma_\zeta)$ in the GMG model & (0, 0.2, 0.02)\\
		Data generation rate & 100Kbps \raisebox{-0.5ex}{\textasciitilde} 600Kbps \\
		Simulation time & 100 s\\
		\bottomrule
	\end{tabular}
\end{table}
  \item PDR: It is defined as the ratio of the number of packets received by all destination nodes to the number of packets sent by all source nodes, i.e.,
      \begin{equation}\label{eq14}
          {\rm {PDR}} = \frac{{\sum N_{i,\rm R} }}{{\sum {N_{j,\rm S}} }},
      \end{equation}
where $\sum N_{i,\rm R}$ and $\sum N_{j,\rm S}$ denote the total number of received packets and the total number of sent packets, respectively. It reflects the transmission reliability of the network using a particular routing protocol.

  \item Network jitter: It is defined as the standard deviation of the E2E latency, i.e.,
      \begin{equation}\label{eq15}
          T_{\rm {Jitter}} = \frac{{\sqrt {\sum {{{(D_{i,\rm{E2E}} - {\overline {D}}_{\rm{E2E}} )}^2}} } }}{{N_{\rm R}  - 1}},
      \end{equation}
      thus it characterizes the network stability in terms of the packet delivery latency.
  \item ROR: It is the ratio of the number of bytes in routing control packets to that in data packets, i.e.,
    \begin{equation}\label{eq16}
    	\mathrm{ROR}=\frac{\sum \textrm{CP}_{\textrm{size},i}}{\sum \textrm{DP}_{\textrm{size},j}},
    \end{equation}
	where $\textrm{CP}_{\textrm{size},i}$ is the size of the $i$th routing control packet, and $\textrm{DP}_{\textrm{size},j}$ is the size of the $j$th data packet. It characterizes how efficient the overhead is used in a routing protocol invoked by the network. The reception of more data packets at the cost of less control packets indicates that the routing protocol is working at higher efficiency.
\end{enumerate}
\subsection{Simulation Results and Discussions}
Simulation parameters regarding the protocol stack are the same as those of \cite{HuCyberPhysical}. Other main simulation parameters, unless explicitly stated otherwise, are given in Table \ref{Table 1}. We analyze the performance of routing protocols in different scenarios, which vary in terms of the network size and traffic load. The network size is denoted by $N_1 \times N_2$, with $N_1$ being the number of UAV groups and $N_2$ being the number of nodes in each UAV group. The traffic load is varied by adjusting the data generation rate.

\begin{figure}[tbp]
	\centering
	\includegraphics[width=3.5in]{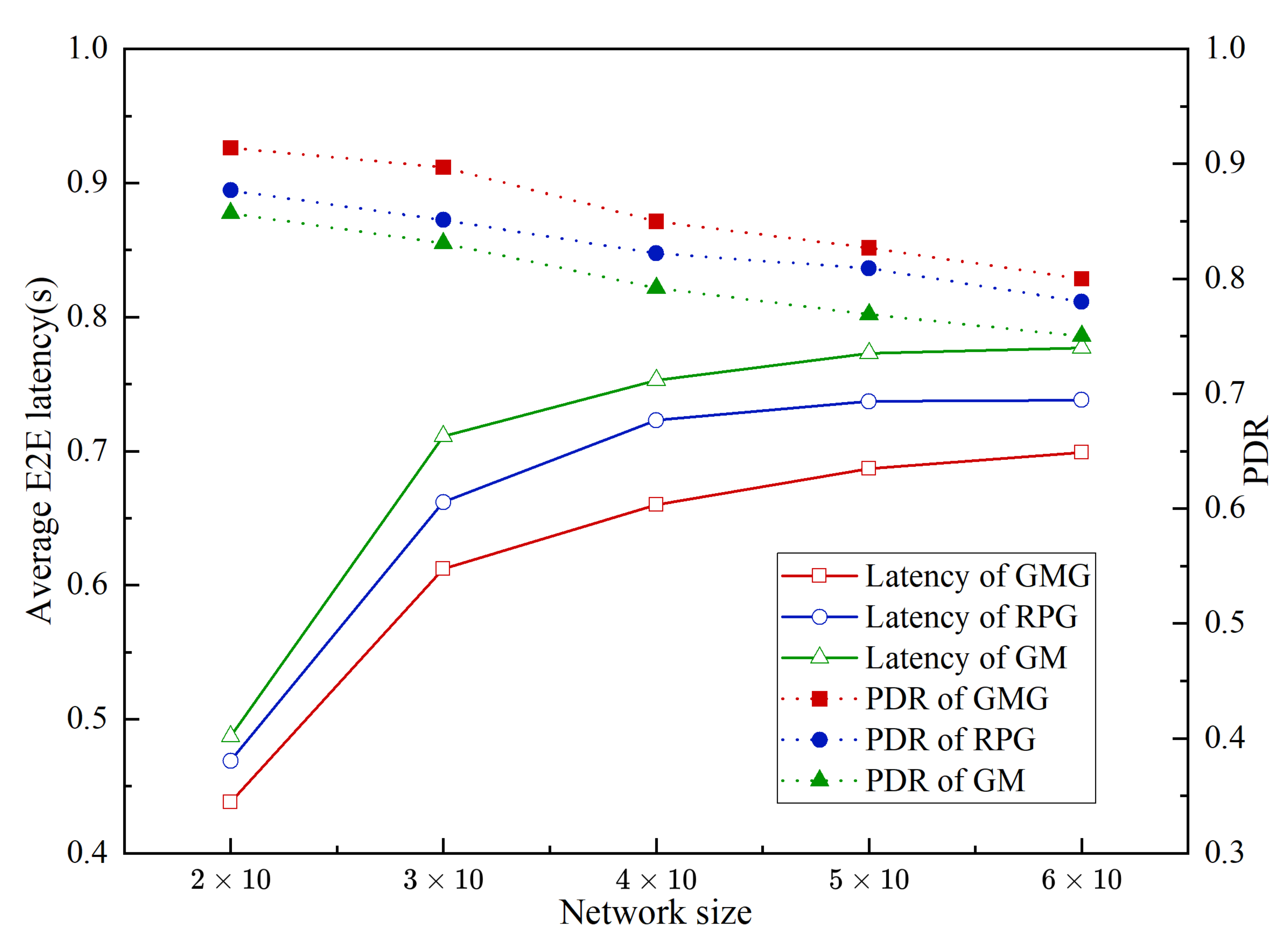}
	\caption{The impact of different mobility models on the PDR and the average E2E latency.}
	\label{Fig5}
\end{figure}

We first evaluate the impact of the GMG, RPG and GM mobility models on the PDR and the average E2E latency based on the AODV routing protocol in Fig.\ref{Fig5}. It is observed that the average E2E latency of the GMG model is lower than that of the RPG and GM model, because there is no sudden stop and turn, and nodes move in coordination in the GMG model. Meanwhile, the GMG model also achieves higher PDR, because it captures the spatio-temporal correlation of node movements in marching formation and the routes selected are more robust.

\begin{figure*}[htbp]
	\centering
	\renewcommand{\thefigure}{6}
	\subfloat[average E2E latency]{
		\includegraphics[width=3.2in]{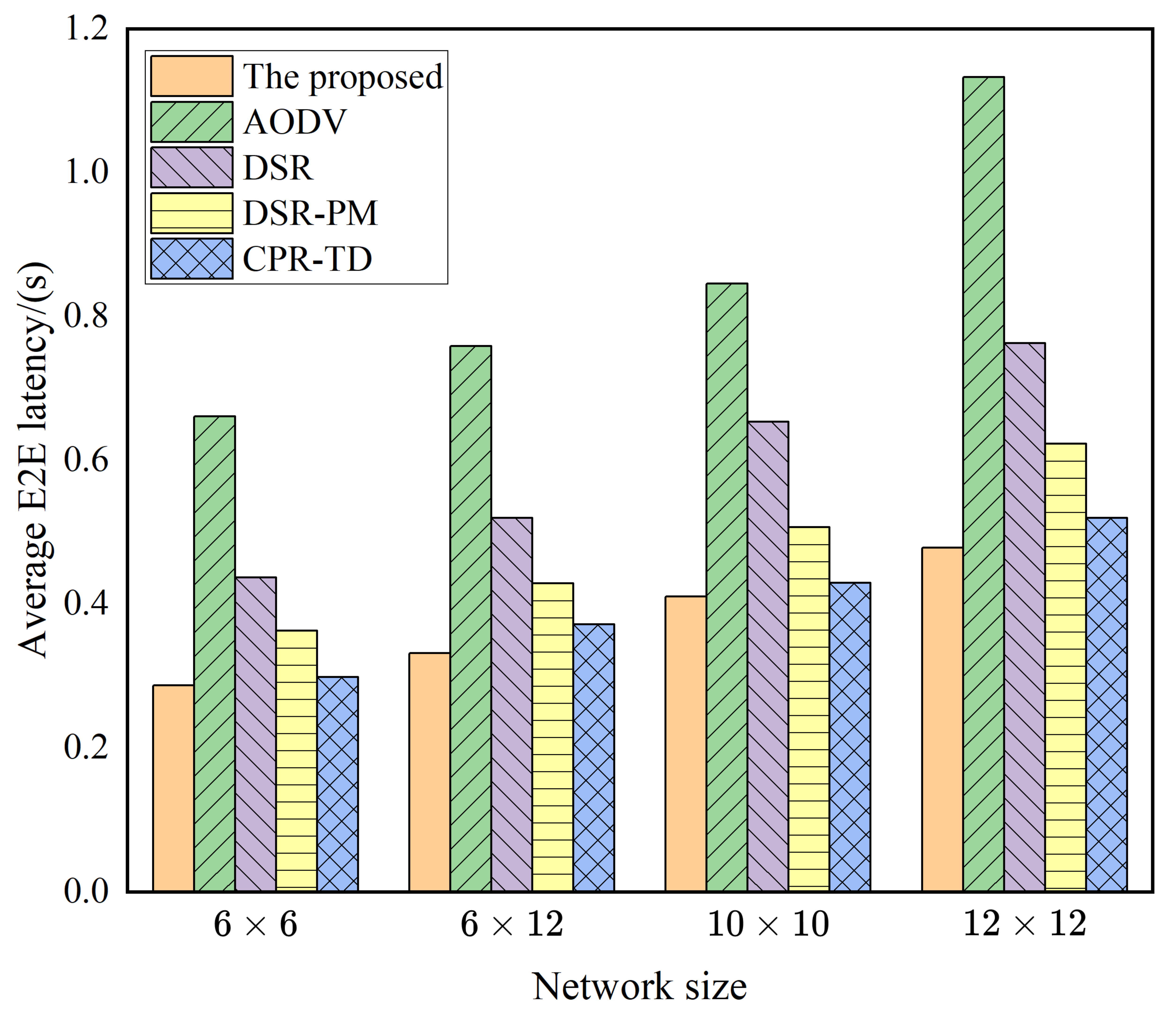}%
		\label{Fig.6(a)}}
	\hspace{0\textwidth} 
	\subfloat[PDR]{
		\includegraphics[width=3.16in]{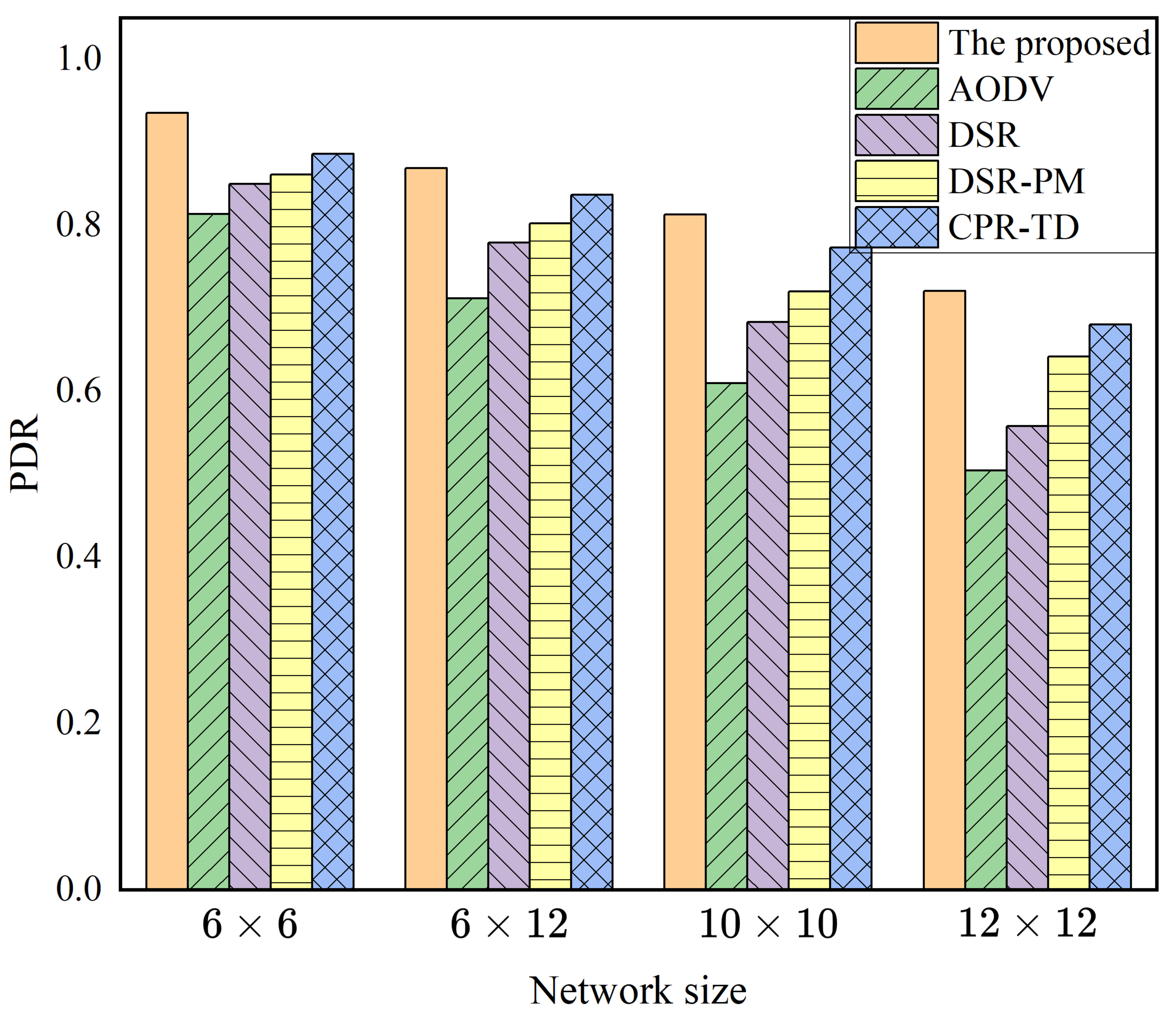}
		\label{Fig.6(b)}}
	\hspace{0\textwidth} 
	\subfloat[network jitter]{
		\includegraphics[width=3.2in]{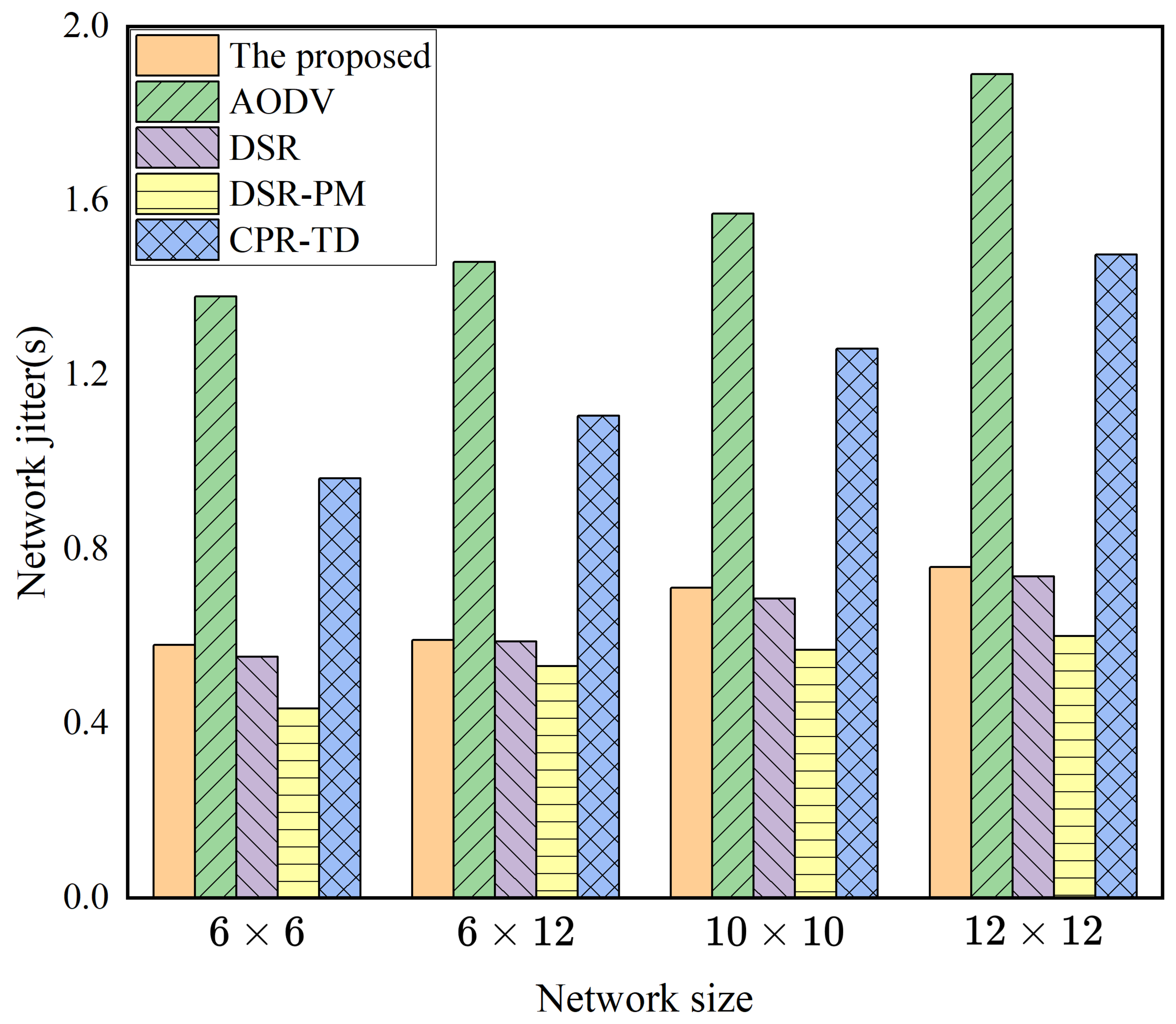}
		\label{Fig.6(c)}}
	\hspace{0\textwidth} 
	\subfloat[ROR]{
		\includegraphics[width=3.26in]{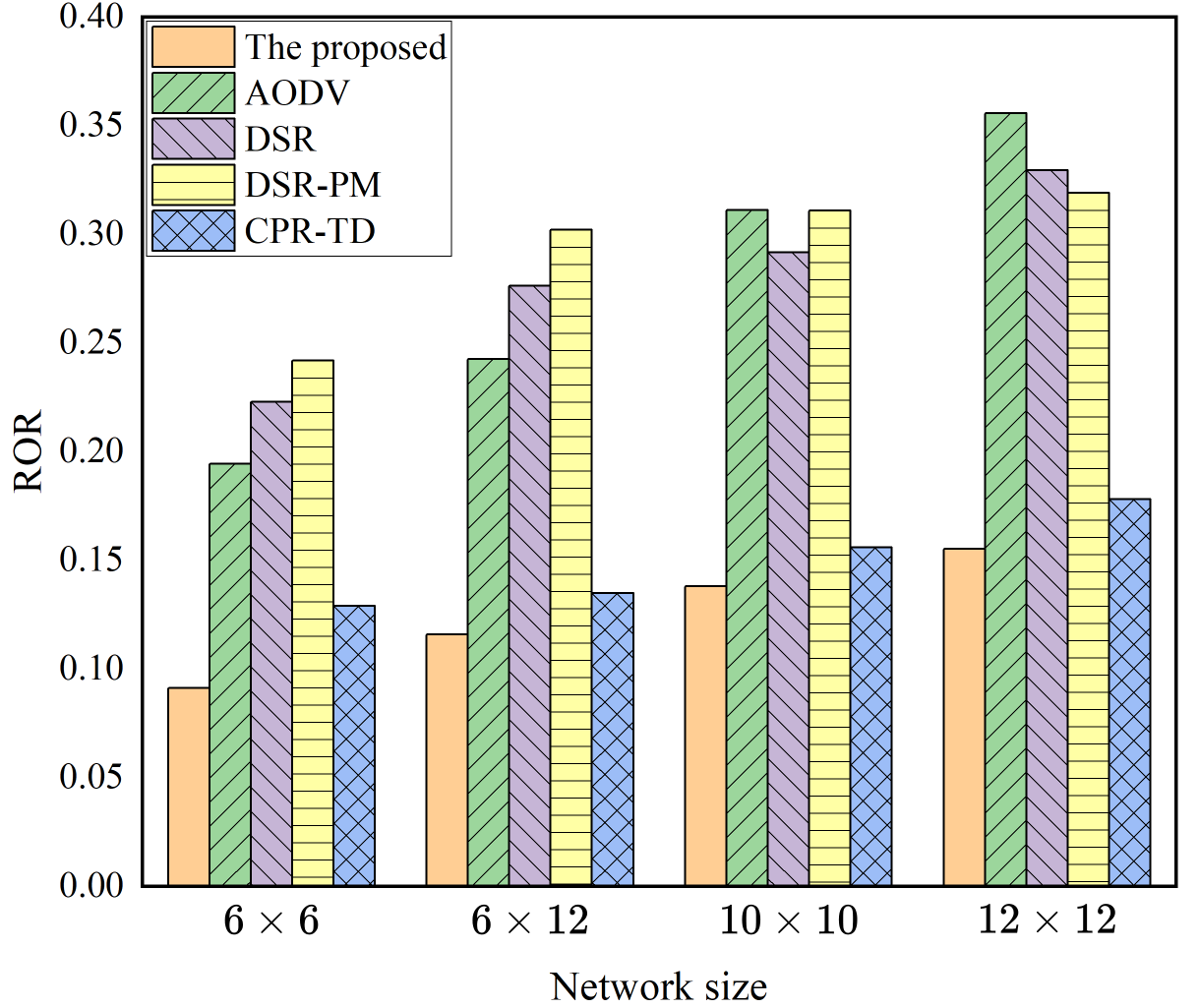}
		\label{Fig.6(d)}}
	\caption{Comparision of the routing performance under different network sizes (data generation rate: 100Kbps).}
	\label{Fig6}
\end{figure*}

Then we consider both widely adopted AODV and DSR, as well as more advanced DSR-PM and CPR-TD for the performance comparison. Fig.\ref{Fig6} compares the routing performance at different network sizes. In Fig.\ref{Fig6}\subref{Fig.6(a)}, we observe that generally the average E2E latency increases as the network scale expands. AODV and DSR, as reactive routing protocols, can quickly find paths in small networks.  However, as network size grows, increased dynamics and RREQ flooding overhead lead to excessive bandwidth use, more collisions, congestion, and high latency, resulting in poor performance. DSR-PM monitors link states and repairs broken links, reducing route reconstruction frequency and lowering average E2E latency. The proposed protocol performs best, followed by CPR-TD. Both leverage mobility model knowledge to optimize path selection and reduce flooding overhead. By exploiting \textit{BC}, our BC-DSR better captures network structure and achieves lower latency than CPR-TD.

In Fig.\ref{Fig6}\subref{Fig.6(b)}, we see that PDR decreases across all protocols as the network size increases, though the extent of this reduction varies by protocol. As the network grows more dynamic, AODV, which stores only one valid route per destination, experiences frequent link breaks, leading to lower PDR. DSR, with multiple stored routes, maintains a relatively higher PDR, but both protocols are unsuitable for FANETs. In contrast, DSR-PM's link monitoring repair mechanism, along with CPR-TD's motion prediction based on trajectory dynamics, enhance link stability and PDR. Additionally, BC-DSR exploit \textit{BC} to create more reliable paths. Since the information about the leader nodes is known \textit{a priori} to other nodes, and data transmission primarily occurs through nodes with high BC. This reduces collisions during route establishment and achieving the highest PDR among the protocols.
\begin{figure*}[htbp]
	\centering
	\renewcommand{\thefigure}{7}
	\subfloat[average E2E latency]{
		\includegraphics[width=3.2in]{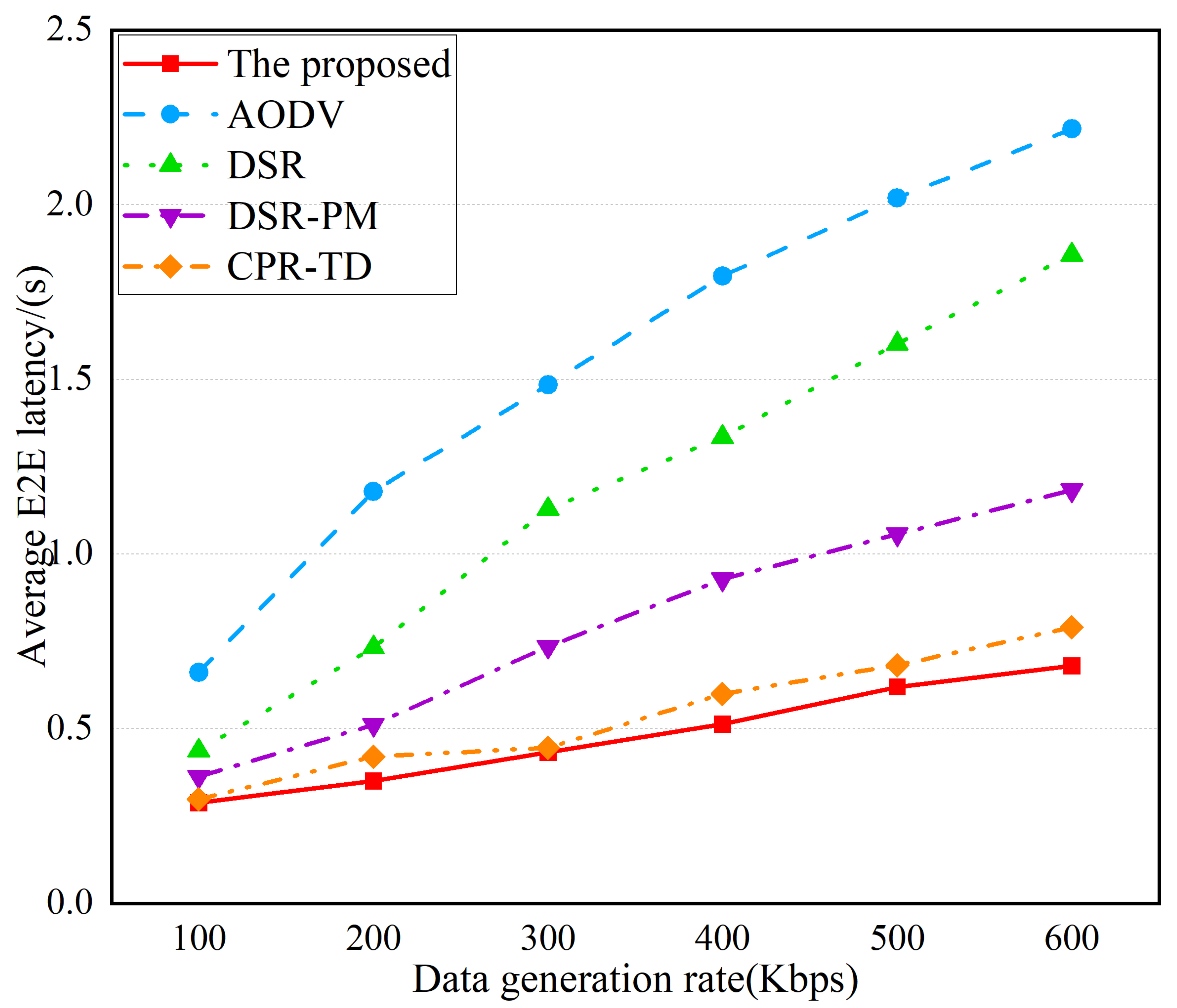}%
		\label{Fig.7(a)}}
	\hspace{0\textwidth} 
	\subfloat[PDR]{
		\includegraphics[width=3.22in]{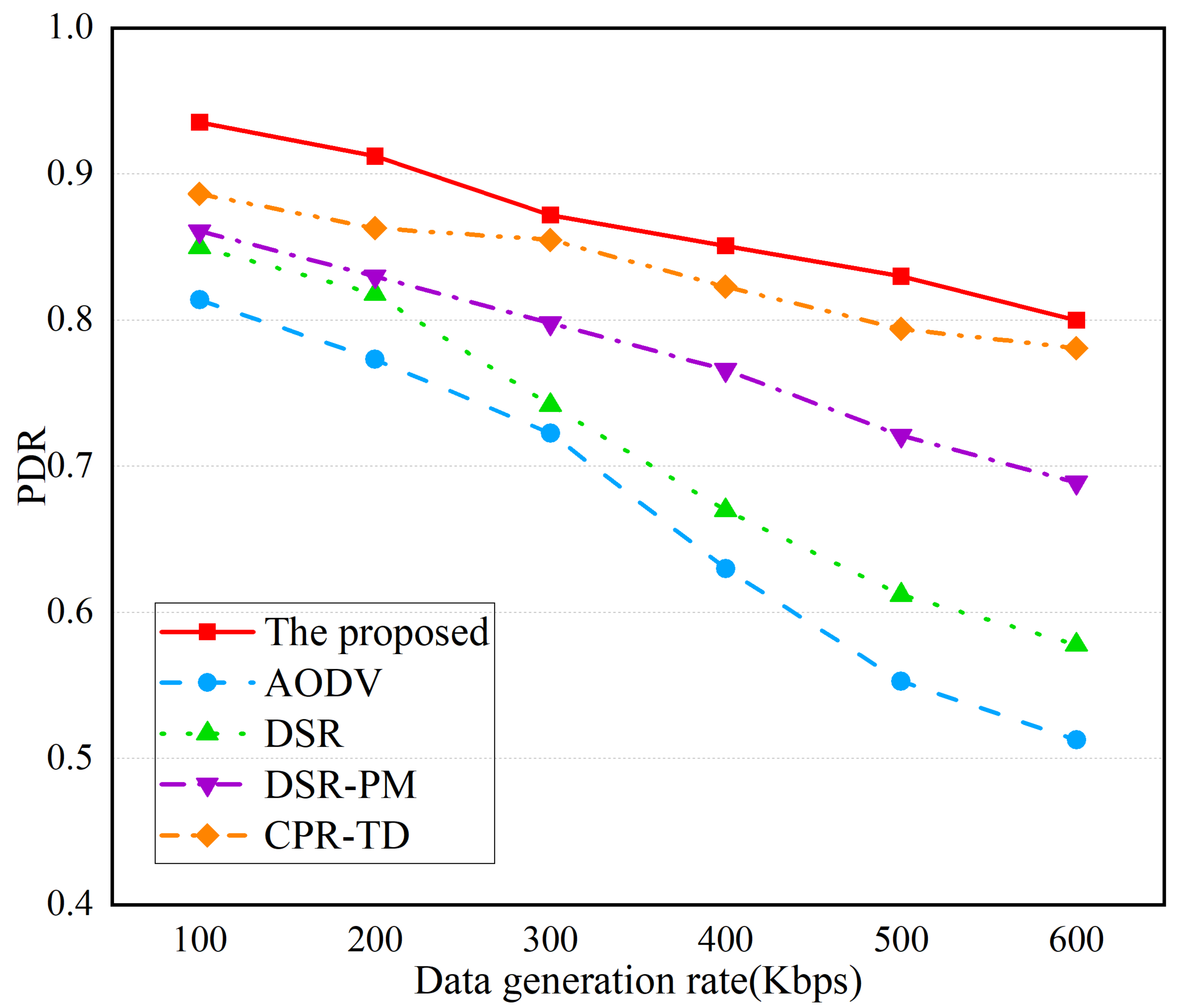}
		\label{Fig.7(b)}}
	\hspace{0\textwidth} 
	\subfloat[network jitter]{
		\includegraphics[width=3.2in]{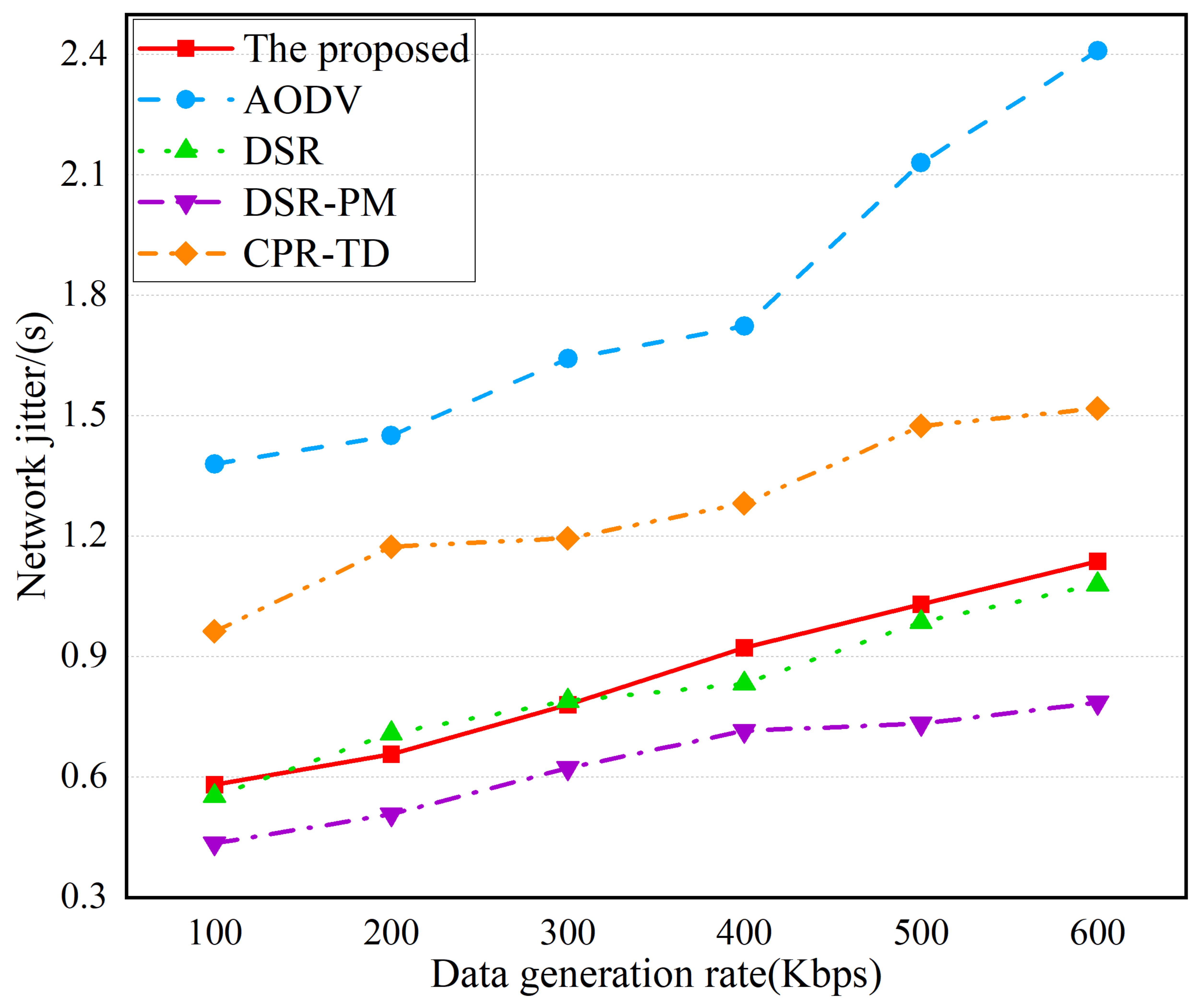}
		\label{Fig.7(c)}}
	\hspace{0\textwidth} 
	\subfloat[ROR]{
		\includegraphics[width=3.3in]{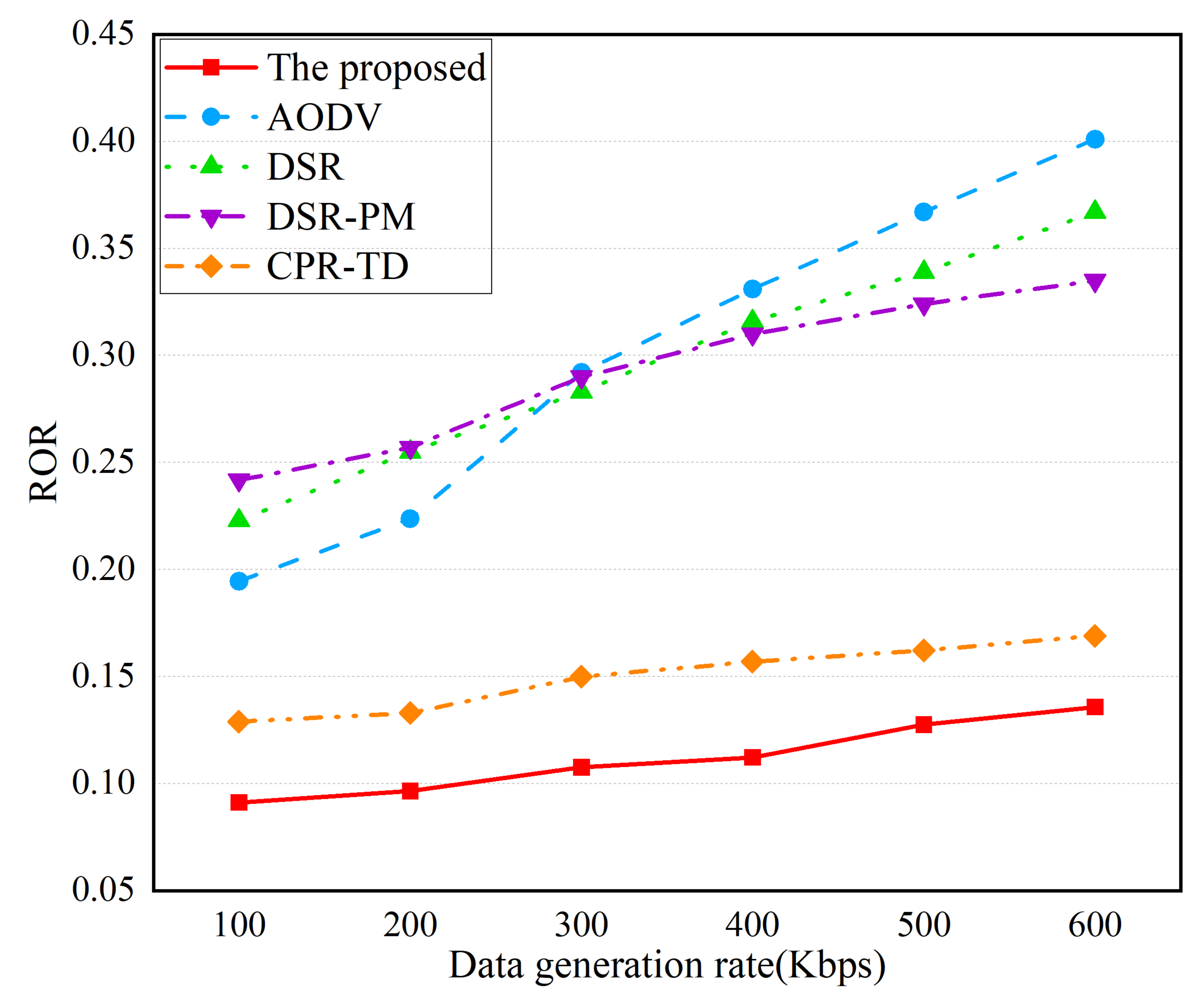}
		\label{Fig.7(d)}}
	\caption{Comparision of the routing performance under different data generation rates (network size: 36).}
	\label{Fig7}
\end{figure*}

In Fig.\ref{Fig6}\subref{Fig.6(c)}, among all the protocols considered, AODV exhibits the highest network jitter, followed by CPR-TD. The network jitter of our BC-DSR is comparable to DSR, but both are slightly higher than DSR-PM. Although not the lowest, BC-DSR still demonstrates competitive performance.

Finally, in Fig.\ref{Fig6}\subref{Fig.6(d)}, the ROR shows a positive correlation with the network size. AODV and DSR experience significant overhead from RREQ flooding due to a large number of control packets, while other protocols have lower overhead. Specifically, our protocol constructs a weighted directed graph based on \textit{BC}, eliminating the need for the route search and maintenance process. Compared to other reactive routing protocols, it avoids the continuous flooding of control packets.
\indent Fig.\ref{Fig7} compares the routing performance under different traffic loads. In Fig.\ref{Fig7}\subref{Fig.7(a)}, as the network traffic load increases, the average E2E latency show varying degrees of growth. AODV and DSR perform reasonably well under low loads but exhibit poor performance under high loads, primarily due to increased collision and congestion. Although DSR-PM introduces the link monitoring repair mechanism that keeps latency low under light loads, it fails to effectively manage the decline in performance under heavy traffic. In contrast, BC-DSR and CPR-TD prevent the flooding of control packets, reducing the probability of collisions, and are able to maintain stability even during high traffic loads. 

In Fig.\ref{Fig7}\subref{Fig.7(b)}, the PDR generally declines as traffic load increases, primarily due to multiple packets competing for limited bandwidth, leading to heightened competition and frequent disruptions, and increased packet loss. For AODV and DSR, the flooding overhead significantly escalates under high traffic conditions, further exacerbating the competition and resulting in a notable reduction in PDR. DSR-PM alleviates these effects by monitoring link quality, while CPR-TD further enhances link stability through trajectory knowledge, leading to an improvement in PDR. In addition, BC-DSR compares the \textit{BC} of each node in a distributed manner to establish more robust routing. Thus, BC-DSR maintains a relatively high PDR even under heavy traffic loads.

In Fig.\ref{Fig7}\subref{Fig.7(c)}, increasing traffic load exacerbates network jitter. Under high load, fluctuations in network state and packet queuing times lead to greater uncertainty in packet arrival times, resulting in increased jitter. Among all the protocols considered, our BC-DSR, while not always the best, consistently maintains a reasonably low level of network jitter.

Finally, in Fig.\ref{Fig7}\subref{Fig.7(d)}, we see that the ROR generally increases with rising traffic load. This is mainly due to heightened competition for network resources, resulting in more conflicts and retransmissions. Consequently, routing protocols like AODV and DSR must handle a greater number of routing control packets. In contrast, BC-DSR effectively manages the increase in ROR by using a simplified routing establishment process and routing cache mechanism.

Jointly observing Fig.\ref{Fig6} and Fig.\ref{Fig7}, we can conclude that our BC-DSR protocol achieves the best overall performance.

\section{Conclusion}
In this paper we first proposed a GMG mobility model for the marching formation scenario of a FANET, where the mobility of nodes both inside a group and between groups is correlated. Then we propose the BC-DSR routing protocol that exploits the BC for route establishment. We compared the PDR, the average E2E latency, the network jitter  and the ROR performance of the proposed BC-DSR protocol under different network sizes and traffic loads with those of the representative benchmark protocols used in FANETs. The ns-3 based simulation results demonstrate that our BC-DSR protocol achieves higher PDR and lower average E2E latency and ROR than baseline protocols, while maintaining a reasonably small network jitter.
The future work will focus on extending our research to better simulate and address real-world challenges.
\ifCLASSOPTIONcaptionsoff
  \newpage
\fi

\bibliographystyle{IEEEtran}
\bibliography{IEEEabrv, reference}

\begin{IEEEbiography}
	[{\includegraphics[width=1in,height=1.25in,clip,keepaspectratio]{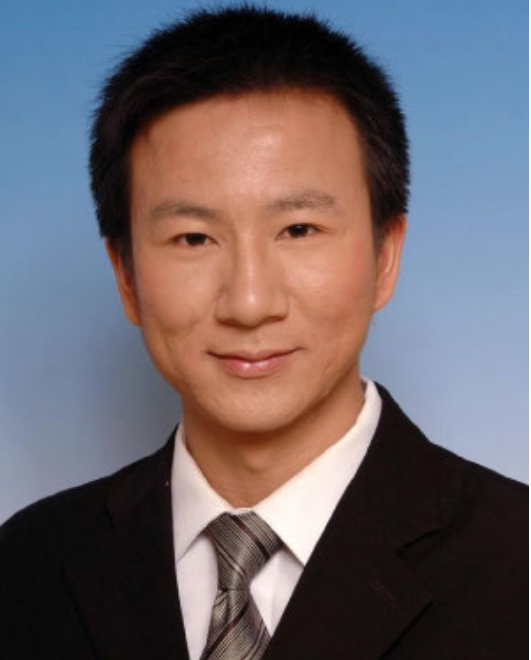}}]
	{Shaoshi Yang}
	(Senior Member, IEEE) 
	received the B.Eng. degree in information engineering from Beijing University of Posts and Telecommunications (BUPT), China, in 2006, and the Ph.D. degree in electronics and electrical engineering from the University of Southampton, U.K., in 2013. From 2008 to 2009, he was a Researcher with Intel Labs China. From 2013 to 2016, he was a Research Fellow with the School of Electronics and Computer Science, University of Southampton. From 2016 to 2018, he was a Principal Engineer with Huawei Technologies Co. Ltd., where he made significant contributions to the products, solutions, and standardization of 5G, wideband IoT, and cloud gaming/VR. He was a Guest Researcher with the Isaac Newton Institute for Mathematical Sciences, University of Cambridge. He is currently a Full Professor with BUPT, a member of the Key Laboratory of Universal Wireless Communications, Ministry of Education, and a deputy director of the Key Laboratory of Mathematics and Information Networks, Ministry of Education. His research expertise includes 5G/5G-A/6G, massive MIMO, mobile ad hoc networks, distributed artificial intelligence, and cloud gaming/VR. He is a Standing Committee Member of the CCF Technical Committee on Distributed Computing and Systems. He received the Dean’s Award for Early Career Research Excellence from the University of Southampton in 2015, the Huawei President Award for Wireless Innovations in 2018, the IEEE TCGCC Best Journal Paper Award in 2019, the IEEE Communications Society Best Survey Paper Award in 2020, the Xiaomi Young Scholars Award in 2023, the CAI Invention and Entrepreneurship Award in 2023, the CIUR Industry-University-Research Cooperation and Innovation Award in 2023, and the First Prize of Beijing Municipal Science and Technology Advancement Award in 2023. He is an Editor of \emph{IEEE Transactions on Communications}, \emph{IEEE Transactions on Vehicular  Technology}, and \emph{Signal Processing} (Elsevier). He was also an Editor of \emph{IEEE Systems Journal} and \emph{IEEE Wireless Communications Letters}. For more details on his research progress, please refer to https://shaoshiyang.weebly.com/.
\end{IEEEbiography}

\begin{IEEEbiography}
	[{\includegraphics[width=1in,height=1.25in,clip,keepaspectratio]{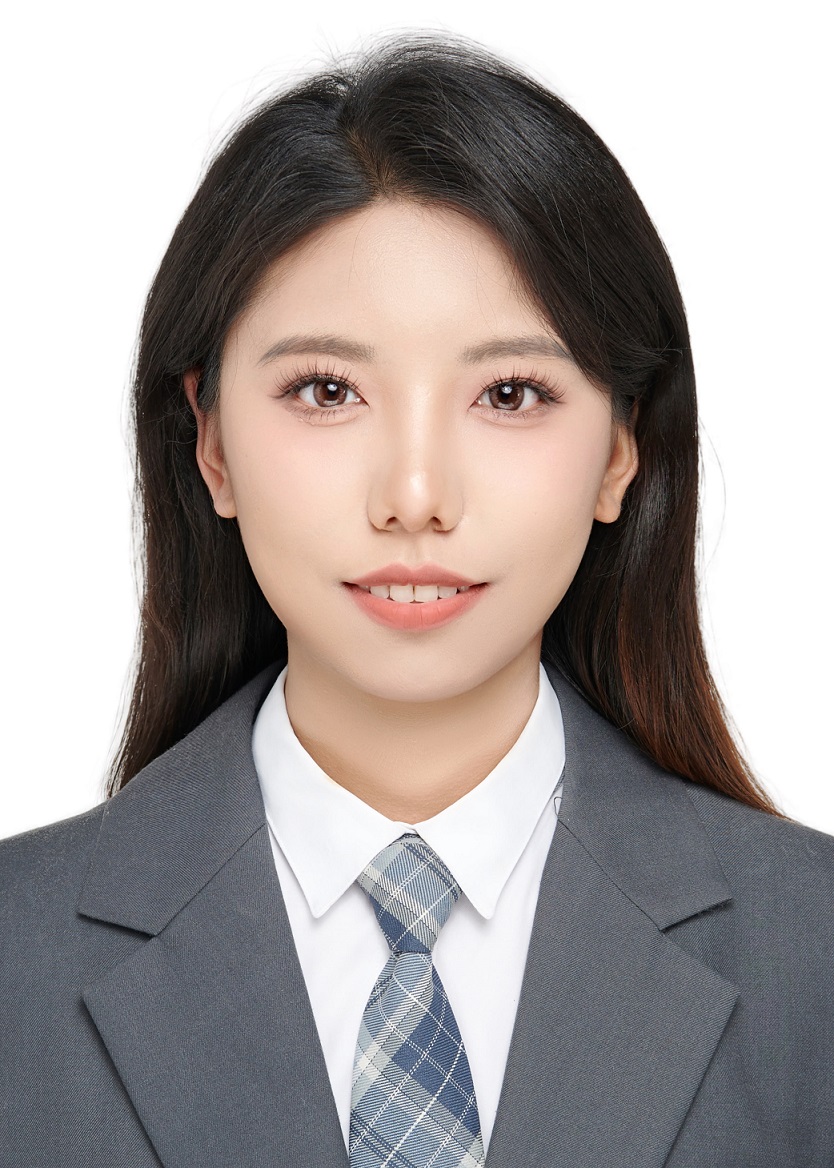}}]
	{Wei Zhao}
	received the B.Eng. degree in communication engineering from Beijing Jiaotong University (BJTU), China, in 2022. She is currently pursuing the M.Eng. degree in information and communication engineering with the School of Information and Communication Engineering, Beijing University of Posts and Telecommunications (BUPT). She is also with the Key Laboratory of Universal Wireless Communications, Ministry of Education, and the Key Laboratory of Mathematics and Information Networks, Ministry of Education. Her current research interests include highly dynamic mobile ad hoc network architecture and protocol stack design, in particular routing protocol design for mobile ad hoc networks.
\end{IEEEbiography}

\begin{IEEEbiography}
	[{\includegraphics[width=1in,height=1.25in,clip,keepaspectratio]{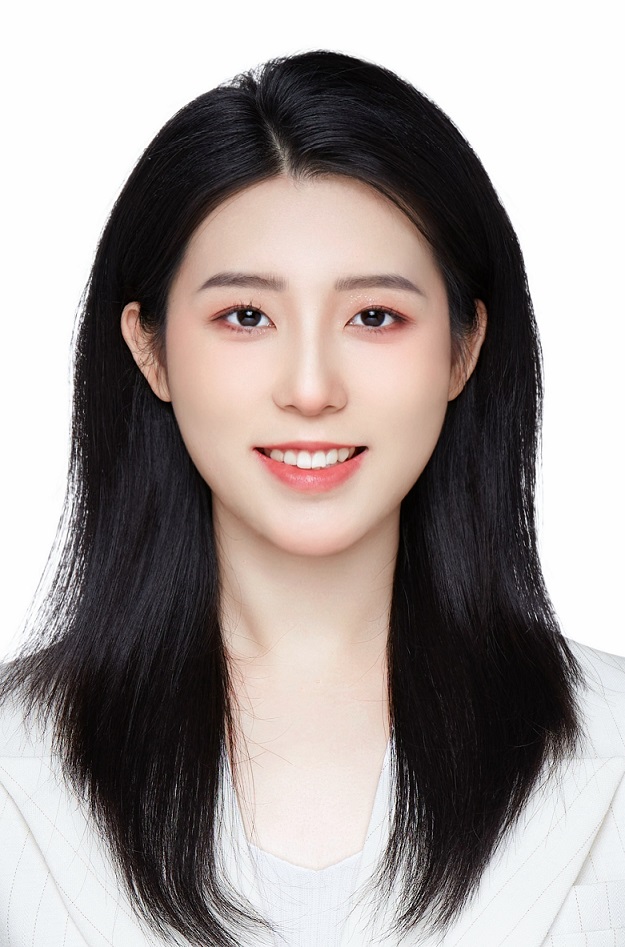}}]
	{Chu-Meng Wang}
	Chu-Meng Wang received the B.Eng. degree in communication engineering from Beijing Jiaotong University, Beijing, China, in 2021, and the M.Eng. degree in information and communication engineering from Beijing University of Posts and Telecommunications, Beijing, China, in 2024. She was also with the Key Laboratory of Universal Wireless Communications, Ministry of Education, and the Key Laboratory of Mathematics and Information Networks, Ministry of Education. Her current research focuses on routing protocol optimization and network performance enhancement for mobile ad hoc networks.
\end{IEEEbiography}

\begin{IEEEbiography}
	[{\includegraphics[width=1in,height=1.25in,clip,keepaspectratio]{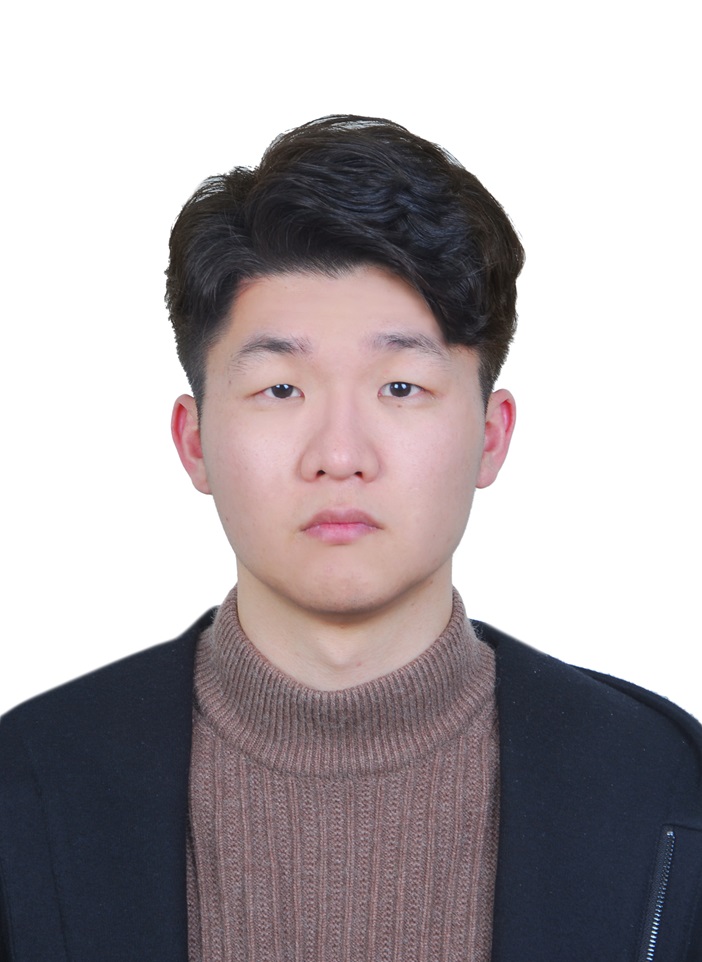}}]
	{Wen-Yu Dong}
	(Student Member, IEEE) 
	received the B.S. degree in electronic and information engineering from Sichuan University (SCU), China, in 2019. He is currently pursuing the Ph.D. degree in information and communication engineering with the School of Information and Communication Engineering, Beijing University of Posts and Telecommunications (BUPT). He is also with the Key Laboratory of Universal Wireless Communications, Ministry of Education, and the Key Laboratory of Mathematics and Information Networks, Ministry of Education. His current research interests include highly dynamic mobile ad hoc network architecture and protocol stack design, routing design for mobile ad hoc networks, and integrated space-air-ground network modeling and performance analysis based on stochastic geometry.
\end{IEEEbiography}

\begin{IEEEbiography}
	{Xiaojie Ju,} photograph and biography not available at the time of publication.
\end{IEEEbiography}

\end{document}